\documentclass[journal]{IEEEtran}
\ifCLASSINFOpdf
\else
\fi
\usepackage{graphicx}
\usepackage{amsmath}
\usepackage{amssymb}
\usepackage{hyperref}

\usepackage{subfigure}

\usepackage{authblk}

\usepackage{cite}
\usepackage{amsmath,amssymb,amsfonts}
\usepackage{algorithmic}
\usepackage{graphicx}
\usepackage{tabularx}
\usepackage{textcomp}
\usepackage{xcolor}

\begin{document}

\title{ReflectGAN: Modeling Vegetation Effects for Soil Carbon Estimation from Satellite Imagery}

\author{Dristi~Datta,~\IEEEmembership{Student Member,~IEEE,}
        Manoranjan~Paul,~\IEEEmembership{Member,~IEEE,}
        Manzur~Murshed,~\IEEEmembership{Member,~IEEE,}
        Shyh~Wei~Teng,~\IEEEmembership{Member,~IEEE,}
        and~Leigh~M.~Schmidtke,~\IEEEmembership{Member,~IEEE}%
        
\thanks{D. Datta and M. Paul are with the School of Computing, Mathematics, and Engineering, Charles Sturt University, Bathurst, NSW 2795, Australia, and also with the Cooperative Research Centre for High Performance Soils, Callaghan, NSW 2308, Australia (e-mail: ddatta@csu.edu.au; mpaul@csu.edu.au).}%
\thanks{M. Murshed is with the School of Information Technology, Deakin University, Burwood, VIC 3125, Australia (e-mail: manzur.murshed@deakin.edu.au).}%
\thanks{S. W. Teng is with the Institute of Innovation, Science and Sustainability, Federation University, Mount Helen, VIC 3350, Australia, and also with the Cooperative Research Centre for High Performance Soils, Callaghan, NSW 2308, Australia (e-mail: s.w.teng@federation.edu.au).}%
\thanks{L. M. Schmidtke is with the Gulbali Institute, Charles Sturt University, Wagga Wagga, NSW 2678, Australia (e-mail: lschmidtke@csu.edu.au).}%
\thanks{Corresponding author: Dristi Datta (e-mail: ddatta@csu.edu.au).}
}

\markboth{Journal of \LaTeX\ Class Files,~Vol.~13, No.~9, September~2014}%
{Shell \MakeLowercase{\textit{et al.}}: Bare Demo of IEEEtran.cls for Journals}



\maketitle

\begin{abstract}

Soil organic carbon (SOC) is a critical indicator of soil health, but its accurate estimation from satellite imagery is hindered in vegetated regions due to spectral contamination from plant cover, which obscures soil reflectance and reduces model reliability. This study proposes the Reflectance Transformation Generative Adversarial Network (ReflectGAN), a novel paired GAN-based framework designed to reconstruct accurate bare soil reflectance from vegetated soil satellite observations. By learning the spectral transformation between vegetated and bare soil reflectance, ReflectGAN facilitates more precise SOC estimation under mixed land cover conditions. Using the LUCAS 2018 dataset and corresponding Landsat 8 imagery, we trained multiple learning-based models on both original and ReflectGAN-reconstructed reflectance inputs. Models trained on ReflectGAN outputs consistently outperformed those using existing vegetation correction methods. For example, the best-performing model (RF) achieved an $R^2$ of 0.54, RMSE of 3.95, and RPD of 2.07 when applied to the ReflectGAN-generated signals, representing a 35\% increase in $R^2$, a 43\% reduction in RMSE, and a 43\% improvement in RPD compared to the best existing method (PMM-SU). The performance of the models with ReflectGAN is also better compared to their counterparts when applied to another dataset, i.e., Sentinel-2 imagery. These findings demonstrate the potential of ReflectGAN to improve SOC estimation accuracy in vegetated landscapes, supporting more reliable soil monitoring.

\end{abstract}


% \fi
\IEEEpeerreviewmaketitle

\section{Introduction}

\IEEEPARstart{S}{oil} organic carbon (SOC) is a fundamental indicator of soil health, influencing agricultural productivity, carbon sequestration, improved soil moisture retention and overall ecosystem sustainability. Accurate estimation of SOC is essential for promoting sustainable agriculture, improving soil management practices, and monitoring environmental changes \cite{fageria2012role, bhattacharya2016review}. Traditional methods for estimating SOC rely on laboratory-based soil analyses, which, although precise, are labor-intensive, costly, and limited in spatial coverage \cite{weil2003estimating, loria2024handheld}. These constraints have led to growing adoption of remote sensing technologies as a scalable and cost-effective alternative to conventional field sampling and laboratory analysis.

Laboratory-based hyperspectral imaging (HSI) provides a powerful tool for SOC estimation by offering high spatial and spectral resolution, enabling detailed analysis of soil properties without the need for destructive sampling \cite{liu2023prediction, li2022remote, datta2022soil}. Numerous studies have validated the effectiveness of HSI in accurately estimating SOC levels \cite{ge2021estimating, datta2022soil}. However, the widespread deployment of HSI is constrained by the high cost of equipment and limited accessibility, making it impractical for large-scale applications. These limitations have prompted interest in more affordable alternatives, such as visual-band multispectral imaging, which can achieve competitive SOC estimation accuracy at substantially lower costs \cite{datta2023comparative, stiglitz2017using}. Still, both HSI and multispectral imaging systems are typically confined to localized areas. To address these spatial constraints, satellite-based remote sensing has emerged as a promising approach for large-scale SOC estimation, offering broader spatial coverage and cost-effective monitoring across diverse landscapes.

Multispectral satellite platforms such as Sentinel-2 (S2) and Landsat 8 (L8) offer broad spatial coverage and public accessibility, making them attractive for large-scale SOC estimation. However, their estimation accuracy often lags behind hyperspectral approaches due to environmental noise, limited spectral resolution, and the need for extensive preprocessing \cite{zhou2021prediction, pande2022prediction, datta2023novel, yuzugullu2024satellite}. To mitigate these challenges, recent studies have integrated supplementary features—such as vegetation indices (VIs), soil indices (SIs), and Tasseled Cap Transformation (TCT) components—alongside raw spectral bands. This feature augmentation has been shown to improve the performance of learning-based models by enhancing their ability to capture the complex variability associated with SOC \cite{datta2023novel}.

Nevertheless, most existing studies have primarily focused on bare soil conditions, where minimal or no vegetation is present. This emphasis stems from the fact that vegetation substantially alters surface reflectance, masking the underlying soil characteristics and complicating SOC estimation. Consequently, models are often trained exclusively on bare soil samples to minimize spectral interference. However, in real-world agricultural and natural landscapes, bare soil exposures are seasonal and limited, underscoring the importance of developing methods that can reliably estimate SOC under vegetated conditions—an area that remains insufficiently explored.

Traditional approaches to vegetation correction in satellite-based SOC estimation include vegetation index-based techniques, spectral unmixing models, and radiative transfer frameworks. VIs, such as NDVI and SAVI, are commonly used to estimate vegetation cover and partially reduce its impact on reflectance measurements \cite{xue2017significant}. While computationally efficient, these methods do not recover the true underlying soil reflectance and often yield poor SOC estimation performance in densely vegetated areas \cite{dos2025integrating}. More physically grounded models, such as PROSAIL \cite{verhoef2007coupled}, simulate canopy–soil reflectance interactions using radiative transfer theory. However, their application is limited by the need for numerous biophysical parameters (e.g., leaf area index, leaf angle, soil moisture), which are difficult to obtain at scale.

Spectral unmixing approaches, such as Spectral Mixture Analysis (SMA) \cite{veraverbeke2012spectral, salih2023spectral} and Adaptive SMA (ASMA) \cite{sun2023global}, attempt to separate soil and vegetation components by estimating fractional abundances of endmembers. While conceptually appealing, these methods assume linear mixing models and often rely on idealized endmembers that may not adequately capture the nonlinear reflectance behavior of real-world landscapes. To address these limitations, several nonlinear unmixing models have been proposed, which are better suited for complex environments where reflectance is inherently nonlinear. These include autoencoder-based networks grounded in the Multilinear Mixing Model, which effectively model complex nonlinear interactions in hyperspectral data \cite{fang2024hyperspectral}, model-based deep autoencoder networks that incorporate prior knowledge of the mixing process to enhance unmixing performance \cite{li2021model}, and interpretable deep learning (DL) methods that account for nonlinearity and endmember variability through a probabilistic variational framework \cite{borsoi2023learning}. Probabilistic extensions like the Probabilistic Mixture Model-based Spectral Unmixing (PMM-SU) \cite{eches2010bayesian, hoidn2024probabilistic} account for uncertainty in unmixing but still treat vegetation as a contaminant to be separated rather than learning to reconstruct a soil-equivalent reflectance.

An alternative to spectral unmixing is to bypass contaminated signals entirely by identifying and excluding vegetation-affected samples. Recent approaches, such as transformer-guided noise detection, aim to filter out anomalous reflectance patterns that deviate from expected soil signals \cite{DATTA2025110406}. Although effective in improving training quality, these techniques reduce sample availability and limit the model's applicability in persistently vegetated regions.

Reconstruction-based methods offer a promising solution by learning to generate bare soil reflectance directly from vegetation-contaminated inputs. Some attention-based DL models have been proposed to model soil–vegetation interactions \cite{datta2024unveiling}, but they do not explicitly reconstruct bare soil reflectance. As a result, residual vegetation signals may persist in the output, reducing interpretability and potentially degrading SOC estimation accuracy. Explicit reconstruction is essential to ensure that the reflectance conforms to bare soil spectral patterns, thereby enhancing the reliability of the SOC model and facilitating comparisons with bare-soil-based studies.

Generative Adversarial Networks (GANs) \cite{goodfellow2014generative} offer a data-driven approach for spectral reconstruction. While Pix2Pix \cite{isola2017image} and CycleGAN \cite{zhu2017unpaired} have been successful in visual tasks, their objective functions typically prioritize image realism and cyclic consistency over spectral fidelity, making them less suitable for reflectance recovery in geophysical applications.

Beyond image-to-image translation, GANs have been widely adopted in remote sensing tasks such as super-resolution \cite{lanaras2018super} and synthetic scene generation \cite{zhu2017deep, bai2023deep}. However, most of these models prioritize visual or spatial details rather than the accurate reconstruction of spectral profiles. StyleGAN and its variants \cite{karras2019style, karras2020analyzing, karras2021alias} excel in generating photorealistic images from latent vectors but lack mechanisms for structured spectral translation. More recently, GANs have shown potential in hyperspectral unmixing, with models like HyperGAN \cite{wang2024pixel}, Wasserstein GAN-based spectral networks \cite{ozkan2018improved}, and adversarial learning frameworks for endmember-abundance separation \cite{tang2020hyperspectral}. A 3D-GAN was also introduced for spectral unmixing in planetary datasets, improving robustness and generalization under data scarcity \cite{suresh2023spectral}. However, these methods primarily estimate abundance fractions or map latent representations rather than reconstructing full reflectance profiles—limiting their utility for direct SOC estimation.

To overcome these limitations, this study proposes ReflectGAN—a novel GAN architecture tailored to reconstruct full bare soil reflectance profiles from vegetation-affected inputs. Unlike traditional unmixing-focused GANs, ReflectGAN does not aim to extract fractional abundances. Instead, it learns a direct mapping between paired vegetated and bare soil reflectance spectra. By integrating residual learning and spectral conditioning, ReflectGAN ensures that the reconstructed spectra retain physical fidelity, enabling robust SOC estimation even in moderately to densely vegetated regions.

ReflectGAN learns the systematic relationship between vegetated and bare soil reflectance using paired training examples, explicitly modeling the structured spectral distortion introduced by vegetation. Rather than treating vegetation as random noise, ReflectGAN conditions its generation on the input vegetated reflectance and incorporates residual learning blocks within the generator to recover the underlying bare soil characteristics. This enables the model to disentangle vegetation interference and reconstruct realistic soil reflectance patterns, even in areas with moderate to dense canopy cover. By generating more accurate bare-soil spectra, ReflectGAN significantly enhances SOC estimation in environments where vegetation has traditionally hindered reliable analysis.

The main contributions of this study are summarized as follows:

\begin{itemize}
    \item \textbf{Development of a GAN architecture for spectral correction:} ReflectGAN introduces a novel paired GAN-based framework specifically designed to transform vegetation-contaminated satellite reflectance into bare soil reflectance. The model is tailored for spectral correction, addressing vegetation-induced spectral distortion, thereby improving the accuracy of SOC estimation from satellite data.

    \item \textbf{Overcoming the limitations of conventional and unmixing-focused GAN models:} Existing GAN architectures, such as Pix2Pix and StyleGAN \cite{isola2017image, karras2019style}, primarily focus on visual realism, often compromising spectral fidelity. Recent GAN-based nonlinear unmixing models mainly aim to estimate abundance fractions or classify spectral mixtures in generative space \cite{wang2024pixel, ozkan2018improved, suresh2023spectral}, rather than reconstructing full, physically consistent reflectance profiles. In contrast, ReflectGAN is explicitly designed to recover complete bare soil reflectance by integrating residual learning to maintain spectral consistency and employing spectral conditioning to ensure that generated reflectance aligns with physically accurate bare soil spectra, thereby preserving physical integrity.

    \item \textbf{Enhanced SOC estimation in vegetated landscapes:} Experiments demonstrate that learning-based models trained on ReflectGAN-reconstructed reflectance consistently outperform those using uncorrected inputs, particularly in vegetated regions. Compared to transformer-based noise filtering approaches \cite{DATTA2025110406}, ReflectGAN supports broader spatial coverage by reconstructing rather than discarding vegetation-affected samples.
\end{itemize}

Through these contributions, ReflectGAN advances the current state of remote sensing-based soil analysis by offering a practical and scalable solution for SOC estimation under real-world, mixed land cover conditions.

The remainder of this paper is organized as follows: Section II describes the dataset preparation, including soil sample selection and satellite data processing. Section III details the proposed ReflectGAN architecture. Section IV outlines the overall SOC estimation method. Section V presents the experimental results, while Section VI provides discussion on key findings, limitations, and future research. Section VII concludes the paper.

\section{Dataset Preparation}

\newcolumntype{L}[1]{>{\centering\let\newline\\\arraybackslash\hspace{0pt}}m{#1}}
\newcolumntype{C}[1]{>{\centering\let\newline\\\arraybackslash\hspace{0pt}}m{#1}}
\newcolumntype{R}[1]{>{\centering\let\newline\\\arraybackslash\hspace{0pt}}m{#1}}
\newcolumntype{S}[1]{>{\centering\let\newline\\\arraybackslash\hspace{0pt}}m{#1}}

\begin{table*}[htp]
%\vspace{-0.5cm}
\caption{Descriptive statistical parameters for the SOC dataset investigated in this study, illustrating the variability and distribution of SOC values used for model training and evaluation.}

%\vspace{-0.3cm} % title of Table
\centering % used for centering table
\begin{tabular}{| L{2.8cm} | L{2.8cm}|L{1.2cm}|L{1.2cm} |L{1.2cm} |L{1.2cm} |L{1.2cm}| L{1.2cm} | L{1.2cm} |}
\hline
\textbf{Soil type} & \textbf{NDVI range} & \textbf{S. No.} & \textbf{Min} & \textbf{Max} & \textbf{Mean} & \textbf{Median} & \textbf{Std.} & \textbf{CV(\%)} \\
\hline
Mixed & $0<\text{NDVI}<0.98$ & 554 & 2.5 & 30 & 14.17 & 12.85 & 6.66 & 47\\
\hline
\end{tabular}
\label{dataset_statistic}
\end{table*}

\subsection{Study Area and Soil Data Collection}

This study utilized the LUCAS (Land Use/Land Cover Area Frame Survey) 2018 topsoil dataset, a standardized soil monitoring initiative coordinated by Eurostat covering EU member states. The dataset comprises 18,984 georeferenced topsoil samples with comprehensive physical and chemical properties, including pH (CaCl\textsubscript{2} and H\textsubscript{2}O), electrical conductivity (EC), SOC, CaCO\textsubscript{3}, phosphorus (P), nitrogen (N), and potassium (K), all sourced from the European Soil Data Centre (ESDAC). Each sample is associated with precise geolocation (\texttt{LUCAS-SOIL-2018.shp}), enabling accurate alignment with satellite imagery for remote sensing applications.

To support reliable SOC estimation from remote sensing, we selected a subset of samples spanning a gradient of vegetation cover, specifically targeting bare soil and barley-vegetated areas. Sample selection was guided by their associated Normalized Difference Vegetation Index (NDVI) values, constrained within the range $0 < \text{NDVI} < 0.98$, capturing conditions from completely bare soil to dense crop cover. This stratified sampling strategy facilitates the focused evaluation of the proposed ReflectGAN framework, which aims to reconstruct accurate soil reflectance by mitigating vegetation-induced spectral distortions in satellite observations.

Table~\ref{dataset_statistic} summarizes the descriptive statistics of SOC in the selected dataset. SOC values exhibit substantial variability, ranging from 2.5 to 30 g/kg, reflecting the diverse soil and land use conditions across the study area. The coefficient of variation (CV) of 47\% underscores the heterogeneity of the dataset—an important factor for assessing the robustness of vegetation correction approaches in SOC estimation pipelines.

\subsection{Landsat 8 Data Collection and Pre-processing}

To integrate satellite-derived spectral information with ground-truth SOC data, we utilized imagery from the L8 Operational Land Imager (OLI) sensor. Images were acquired via the USGS Earth Explorer portal (\url{https://earthexplorer.usgs.gov/}) by spatially matching each soil sample’s coordinates. Temporal filtering was applied to ensure consistency with field measurements, restricting image acquisition to within $\pm$15 days of the sampling date. Additionally, only images with less than 10\% cloud cover were selected to minimize atmospheric interference and ensure reflectance reliability.

Among the available L8 bands, we focused on Bands 1 through 7 (B1–B7), which have demonstrated relevance in soil and vegetation studies \cite{loveland2016landsat, datta2023novel}. Thermal and atmospheric bands (e.g., B10–B11) were excluded to retain only spectrally informative features for SOC estimation \cite{barsi2014landsat}.

The satellite images underwent a multi-step preprocessing workflow. First, radiometric calibration converted digital number (DN) values into top-of-atmosphere reflectance, incorporating solar angle and scene metadata \cite{thorne1997radiometric}. Atmospheric correction was then applied using the FLAASH (Fast Line-of-sight Atmospheric Analysis of Spectral Hypercubes) algorithm \cite{gao2009atmospheric}, correcting for scattering and absorption to produce surface reflectance.

To enhance spatial resolution, Gram-Schmidt pansharpening was performed by fusing the 30-meter multispectral bands with the 15-meter panchromatic band (B8) \cite{maurer2013pan}. This step preserved spectral fidelity while improving the spatial detail necessary for capturing subtle soil variability.

Preprocessing tasks were executed in ENVI (v5.6.1), while georeferencing and reflectance extraction were performed using ArcGIS Pro (v2.8.0). Final surface reflectance values from the selected seven bands were extracted at each sample’s precise geolocation and compiled into an Excel spreadsheet for subsequent analysis.

\subsection{Landsat 8 Image Transformation and Vegetation Indices}

To enhance the spectral feature space and improve SOC modeling, a series of vegetation and soil indices was derived from the reflectance values of the seven selected L8 bands. VIs are widely used to assess canopy density and photosynthetic activity, while SIs help characterize surface attributes such as texture, moisture, and color—factors closely associated with SOC variability.

In this study, we calculated several commonly used VIs, including the Ratio Vegetation Index (RVI), NDVI, Green NDVI (GNDVI), Enhanced Vegetation Index (EVI), Soil-Adjusted Vegetation Index (SAVI), and Modified SAVI (MSAVI). Additionally, a set of soil-related indices was derived, such as the Brightness Index (BI), Salinity Index (SI), Color Index (CI), Dry Soil Index (DSI), and Dry Vegetation Index (DVI). These indices were computed using established spectral band formulas described in prior literature \cite{gitelson1996use, qi1994modified, dehni2012remote, datta2023novel}.

Furthermore, the TCT was applied to extract three orthogonal components—brightness, greenness, and wetness—which summarize dominant spectral trends in the landscape. TCT is particularly effective for capturing joint soil–vegetation dynamics and improving the interpretability of multispectral data. The mathematical formulation and implementation details of TCT are described in \cite{baig2014derivation}.

The final feature set integrates raw reflectance values from L8 with the derived VIs, SIs, and TCT components, yielding a comprehensive and biophysically informed representation of the surface environment. This enriched dataset provides a robust foundation for learning-based SOC estimation across heterogeneous land cover conditions.

\subsection{Paired Dataset Preparation for ReflectGAN}

To train ReflectGAN, a paired dataset of vegetated and corresponding bare soil reflectance samples was constructed. Vegetated samples were selected with NDVI~$>$~0.2 to represent significant vegetation cover, while bare soil samples were identified with NDVI~$<$~0.2 to ensure minimal vegetation influence. Each vegetated sample was paired with its closest bare soil counterpart by minimizing the Euclidean distance between their geographic coordinates, under the assumption that SOC levels do not vary abruptly within small spatial extents \cite{mulla2001soil}. This matching ensures that paired samples share similar soil properties apart from the influence of vegetation cover.

In cases where multiple bare soil samples were found in close proximity to a vegetated sample, we computed the average reflectance of the nearest candidates to serve as a robust reference. This averaging approach reduces the influence of local noise or outliers and ensures that the resulting bare soil target accurately represents typical reflectance in the surrounding area.

The final paired dataset consists of input–target reflectance pairs, where the input corresponds to the multispectral reflectance of a vegetated soil sample and the target is the aggregated bare soil reflectance. These pairs were used to train ReflectGAN in a supervised learning setting, enabling the model to learn the structured transformation from vegetation-contaminated observations to vegetation-suppressed soil reflectance.

\section{Proposed ReflectGAN}

\subsection{Proposed ReflectGAN Architecture}

ReflectGAN is a specialized GAN architecture developed to accurately recover bare soil reflectance from vegetation-contaminated satellite imagery, thereby enhancing the estimation of SOC in partially or fully vegetated regions. The model leverages a conditional GAN structure, where the generator is enhanced with residual learning blocks to support detailed and effective spectral transformation from vegetated to bare soil conditions. This architecture is tailored to capture the complex, non-linear relationships inherent in high-dimensional spectral data, effectively addressing the spectral interference caused by vegetation cover. 

By training on paired samples of vegetation-affected and corresponding bare soil reflectance, ReflectGAN effectively isolates and removes vegetation signals, recovering accurate soil-specific spectral information. This capability not only enhances the interpretability of remote sensing data but also contributes significantly to advancing soil monitoring, sustainable land management, and precision agriculture.

\begin{figure*}
  \centering
  \includegraphics[width=7in]{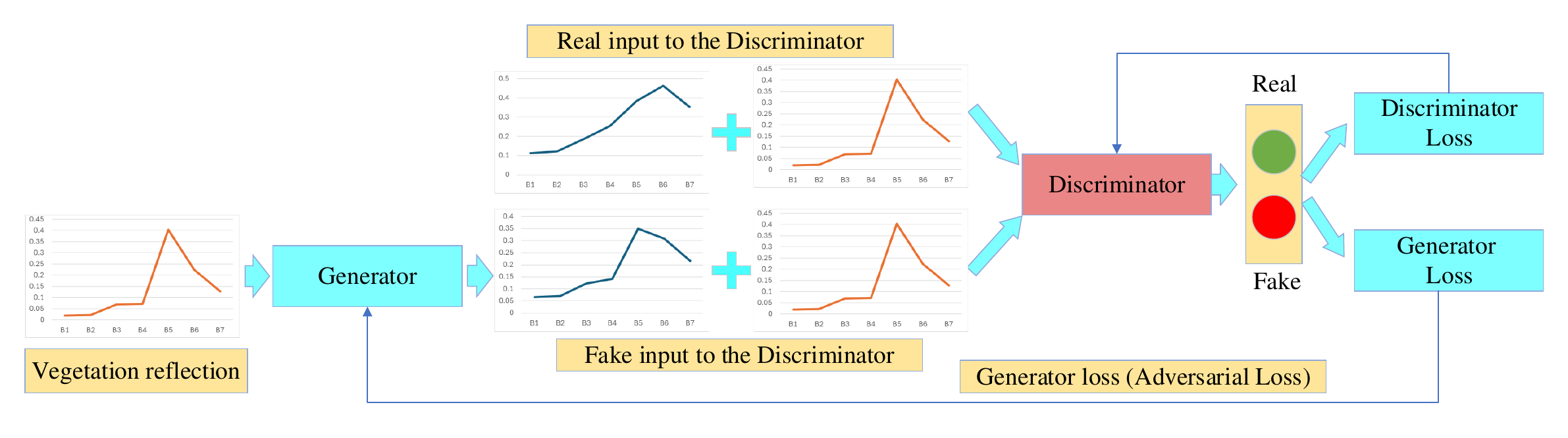}
\caption{The proposed ReflectGAN architecture to illustrate the interaction between the Generator and Discriminator, where the Generator transforms vegetated soil reflectance into bare soil-like spectra, and the Discriminator evaluates their authenticity.}  \label{GAN}
\end{figure*}

The architecture of the proposed ReflectGAN consists of two primary components: a generator and a discriminator, each playing a distinct role within the adversarial learning framework illustrated in Fig.~\ref{GAN}. The generator is responsible for translating vegetated soil reflectance into bare soil spectral signatures, while the discriminator aims to distinguish between real bare soil reflectance and the generator’s synthetic outputs. This adversarial setup enables the generator to improve progressively through feedback, enhancing its ability to reconstruct soil-specific spectral patterns from vegetation-contaminated inputs.

The generator in ReflectGAN is designed to translate vegetated soil reflectance into bare soil spectra, conditioned on the input reflectance. It leverages residual learning blocks to preserve high-level features and improve spectral transformation, allowing the model to effectively learn and represent complex relationships between vegetated and bare soil observations. The Discriminator, on the other hand, is trained to distinguish real bare soil spectra from the generator’s synthetic outputs, while being conditioned on the vegetated inputs. This adversarial setup encourages the generator to produce increasingly realistic and context-aware soil reflectance.

ReflectGAN builds upon a conditional GAN framework and introduces residual learning to enhance its capability for vegetation-to-bare soil reflectance conversion. A key innovation of ReflectGAN is the incorporation of spectral conditioning, where the generation process is explicitly guided by input vegetated reflectance to maintain physical consistency. Unlike standard GANs, which typically generate data without leveraging spectral context, ReflectGAN's spectral conditioning enables more targeted and meaningful generation of bare soil spectra. This novel approach significantly improves the preservation of spectral fidelity in vegetation-contaminated satellite data, making ReflectGAN particularly suitable for SOC estimation under mixed land cover conditions. The standard GAN objective is defined as follows:

\begin{equation}
\begin{split}
\min_{G} \max_{D} V(D, G) = & \mathbb{E}_{x \sim p_{\text{data}}(x)}[\log D(x)] \\
& + \mathbb{E}_{z \sim p_{z}(z)}[\log(1 - D(G(z)))]
\end{split}
\end{equation}
Here, $x$ denotes real samples from the true data distribution $p_{\text{data}}$, and $z$ is a noise vector sampled from a prior distribution $p_z(z)$. The Discriminator $D$ learns to maximize this objective by correctly identifying real and fake data, while the Generator $G$ attempts to minimize it by producing synthetic data indistinguishable from real samples.

ReflectGAN extends this formulation by incorporating conditional information in both the Generator and Discriminator. The modified objective is:

\begin{equation}
\begin{split}
\min_{G} \max_{D} V(D, G) = & \mathbb{E}_{x \sim p_{\text{data}}(x)}[\log D(x|y)] \\
& + \mathbb{E}_{z \sim p_{z}(z)}[\log(1 - D(G(z|y)))]
\end{split}
\end{equation}

In this formulation, $y$ represents the vegetated soil reflectance used as the conditioning variable. The Generator $G$ produces synthetic bare soil spectra $G(z|y)$ based on input $y$, while the Discriminator $D$ assesses whether a given spectrum is a realistic bare soil sample given the vegetated input $y$.

This conditional formulation enables ReflectGAN to generate outputs that are not only spectrally realistic but also contextually accurate. By incorporating residual learning and spectral conditioning, ReflectGAN effectively mitigates vegetation-induced noise and improves the fidelity of soil reflectance reconstruction. As a result, it establishes a robust solution for SOC estimation in vegetated landscapes.

\subsection{Calculation of Loss in ReflectGAN}

In the ReflectGAN framework, adversarial training between the Generator ($G$) and the Discriminator ($D$) is governed by well-defined loss functions that guide the model toward accurately reconstructing bare soil reflectance from vegetation-contaminated inputs. These loss functions ensure both the Generator’s ability to synthesize realistic spectra and the Discriminator’s capacity to distinguish real from synthetic data, based on inputs.

\textbf{Discriminator Loss:} The Discriminator is trained to differentiate between real bare soil spectral signatures and those generated by the Generator, both conditioned on vegetated soil inputs. Its loss function consists of two components:

\textit{Real Data Loss ($\text{Loss}_{D_{\text{real}}}$)}: This term penalizes incorrect classification of real bare soil data. Ideally, $D$ should output values close to 1 for real samples:

\begin{equation}
\text{Loss}_{D_{\text{real}}} = -\frac{1}{N} \sum \log(D(x|y))
\end{equation}

\textit{Generated Data Loss ($\text{Loss}_{D_{\text{fake}}}$)}: This term penalizes the Discriminator when it incorrectly classifies synthetic data as real. The Generator's output $G(z|y)$ is expected to be classified close to 0:

\begin{equation}
\text{Loss}_{D_{\text{fake}}} = -\frac{1}{N} \sum \log(1 - D(G(z|y)))
\end{equation}

The total Discriminator loss is the sum of both components:

\begin{equation}
\text{Loss}_D = \text{Loss}_{D_{\text{real}}} + \text{Loss}_{D_{\text{fake}}}
\end{equation}

\textbf{Generator Loss:} The Generator is trained to fool the Discriminator by generating bare soil reflectance that appears real. Its loss encourages $D$ to classify synthetic data as authentic:

\begin{equation}
\text{Loss}_G = -\frac{1}{N} \sum \log(D(G(z|y)))
\end{equation}

Here, $x$ denotes real bare soil reflectance, $y$ represents vegetated soil input (the conditioning variable), $z$ is the noise vector, and $N$ is the batch size.

By incorporating the conditioning variable $y$ into both networks, ReflectGAN ensures that synthetic outputs are not only realistic but also contextually aligned with the input vegetated reflectance. This conditional structure significantly enhances the model’s ability to recover soil-specific signals and suppress vegetation-induced noise.

In summary, these adversarial losses enable ReflectGAN to learn complex, nonlinear mappings between vegetated and bare soil spectral signatures. Iterative optimization of $\text{Loss}_D$ and $\text{Loss}_G$ drives the Generator toward producing physically meaningful, high-fidelity reflectance outputs, ultimately improving soil organic carbon estimation from satellite imagery.

\subsection{Detailed Architecture Overview}

Building upon the previously outlined framework, this section presents a detailed description of the generator and discriminator architectures within ReflectGAN. A close examination of the internal design choices elucidates how each component contributes to the core objective of reconstructing realistic, environmentally consistent bare soil spectral signatures from vegetated soil inputs.

\begin{figure*}
  \centering
  \includegraphics[width=7in]{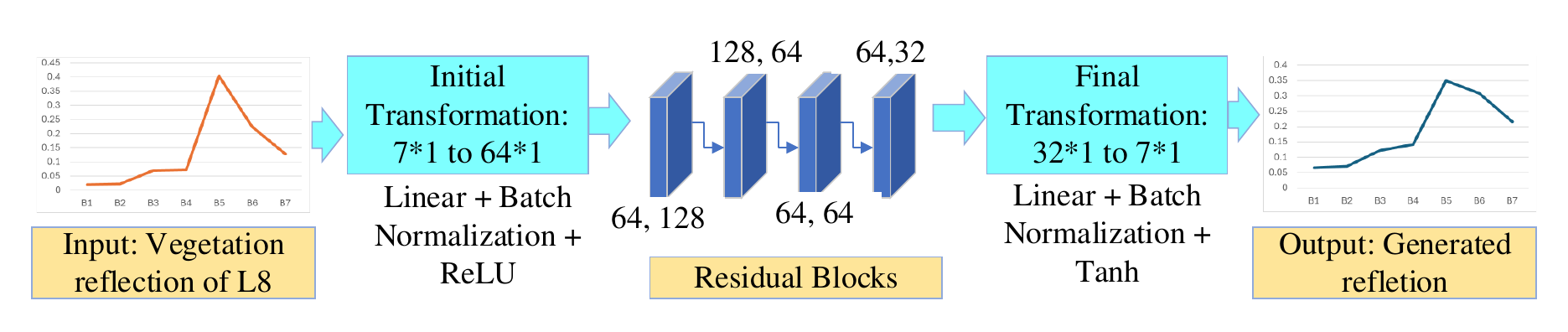}
  \caption{ Schematic Diagram of the Proposed ReflectGAN Generator, illustrating the flow from input vegetated soil spectral signatures through the series of residual blocks to the output of generated bare soil spectral signatures.}\label{GAN_generator}
\end{figure*}

\begin{figure*}
  \centering
  \includegraphics[width=7in]{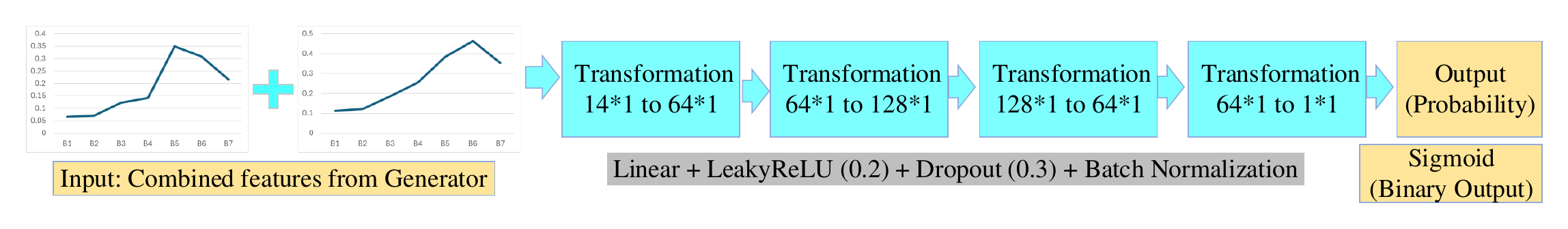}
  \caption{Schematic Diagram of the Proposed ReflectGAN Discriminator, showcasing the sequence of processing layers from concatenated vegetated and bare soil spectral signature inputs to the binary classification output.}\label{GAN_discriminator}
\end{figure*}

\subsubsection{Generator}

The Generator in ReflectGAN is designed to transform vegetated soil spectral signatures into bare soil spectral signatures using L8 imagery. It follows a deep neural network structure built around a series of residual blocks, enabling effective feature learning and spectral translation. A schematic overview of the Generator architecture is shown in Fig.~\ref{GAN_generator}.

Initial Linear Transformation: The transformation begins with an initial linear layer that projects the 7-dimensional input (representing the seven L8 bands) into a 64-dimensional feature space. This expansion allows the network to capture richer spectral representations. Batch normalization is applied to stabilize the training process, followed by a Rectified Linear Unit (ReLU) activation function to introduce non-linearity. Mathematically, this step can be described as:

\begin{equation}
x_1 = \text{ReLU}\left(\text{BN}\left(W_0 x + b_0\right)\right)
\end{equation} 
where $W_0 \in \mathbb{R}^{64 \times 7}$ and $b_0 \in \mathbb{R}^{64}$ are the weights and biases of the linear transformation, $\text{BN}$ denotes batch normalization, and $\text{ReLU}$ represents the activation function.

Residual Blocks: At the core of the Generator are multiple Residual Blocks. Each block applies two linear transformations with batch normalization and ReLU activation, followed by an identity skip connection that adds the block's input to its output. This structure enables the model to refine feature representations while preserving essential information. The residual learning within each block is mathematically defined as:

\begin{equation}
\begin{split}
x_{i+1} = & x_i + \text{ReLU}\left(\text{BN}\left(W_{i1} \cdot \text{ReLU}\left(\text{BN}\left(W_{i0} x_i + b_{i0}\right)\right) + b_{i1}\right)\right)
\end{split}
\end{equation}
where $W_{i0}, W_{i1}$ and $b_{i0}, b_{i1}$ are the weights and biases of the linear layers in the $i$-th residual block.

The feature dimensionality within the residual blocks is strategically adjusted: the network initially expands the features from 64 to 128 dimensions for enhanced representation, maintains or compresses to 64 dimensions in intermediate stages, and eventually reduces the features to 32 dimensions. This careful balance between expansion and compression optimizes the network’s ability to capture both fine and broad spectral patterns.

Final Transformation to Output: After passing through the sequence of residual blocks, the Generator concludes with a final linear layer that reduces the feature space from 32 dimensions back to 7 dimensions, aligning with the original L8 spectral band structure. A $\tanh$ activation function is applied at the output layer to normalize the predicted bare soil reflectance values within a bounded range. The final transformation can be expressed as:

\begin{equation}
G(x) = \tanh\left(W_f x_N + b_f\right)
\end{equation}
where $W_f \in \mathbb{R}^{7 \times 32}$ and $b_f \in \mathbb{R}^{7}$ are the weights and biases of the output layer.

Through initial feature expansion, deep residual learning, and final dimensionality adjustment, the Generator effectively reconstructs bare soil reflectance signatures from vegetation-affected inputs. This architecture enables ReflectGAN to learn complex non-linear mappings and contributes to advancing remote sensing-based soil property estimation.

\subsubsection{Discriminator}

The Discriminator in the proposed ReflectGAN model plays a critical role in the adversarial learning framework by distinguishing real bare soil spectral signatures from generated ones, as shown in Fig.~\ref{GAN_discriminator}. Acting as a binary classifier, it challenges the Generator to produce increasingly realistic outputs by evaluating the authenticity of concatenated vegetated and bare soil spectral signatures.

Input Processing: The Discriminator receives two inputs: the vegetated soil spectral signature and the corresponding bare soil signature, which may be real or generated. These two 7-dimensional vectors are concatenated into a single 14-dimensional input:

\begin{equation}
z = \text{concat}(x, y)
\end{equation}
where $x$ represents the vegetated soil spectrum and $y$ denotes the bare soil spectrum. This concatenation enables the Discriminator to jointly analyze both inputs for subtle differences.

Sequential Processing Layers: The concatenated input passes through a sequence of linear transformations and non-linear activations:

\begin{itemize}
    \item \textit{Linear Expansion:} A fully connected layer projects the 14-dimensional input to a 64-dimensional space:
    \begin{equation}
    z_1 = W_1 z + b_1
    \end{equation}
    where $W_1 \in \mathbb{R}^{64 \times 14}$ and $b_1 \in \mathbb{R}^{64}$ are the layer's weights and biases.
    
    \item \textit{LeakyReLU Activation:} A LeakyReLU activation with a negative slope of 0.2 is applied to introduce non-linearity and maintain gradient flow:
    \begin{equation}
    z_2 = \text{LeakyReLU}(z_1, 0.2)
    \end{equation}
    
    \item \textit{Dropout:} A dropout layer with a probability of 0.3 is incorporated to prevent overfitting:
    \begin{equation}
    z_3 = \text{Dropout}(z_2, 0.3)
    \end{equation}
    
    \item \textit{Batch Normalization:} Batch normalization is used to stabilize training and improve convergence:
    \begin{equation}
    z_4 = \text{BN}(z_3)
    \end{equation}
    
    \item \textit{Further Linear Layers:} Additional layers are used to compress and refine the features (64 to 128 dimensions, then 128 to 64 dimensions), each followed by LeakyReLU activations and dropout regularization:
    \begin{equation}
    z_N = \text{LeakyReLU}(\dots(\text{Linear}(\dots)))
    \end{equation}
\end{itemize}

Final Classification: After processing, a final linear layer reduces the feature space to a single scalar, followed by a sigmoid activation to output a probability score between 0 and 1:

\begin{equation}
D(x, y) = \sigma(W_f z_N + b_f)
\end{equation}
where $W_f \in \mathbb{R}^{1 \times 64}$ and $b_f \in \mathbb{R}$ are the final weights and biases, and $\sigma$ denotes the sigmoid function.

The sequential architecture of the Discriminator progressively refines the feature representations, enabling accurate classification between real and generated bare soil reflections. By effectively conditioning on vegetated soil inputs, the Discriminator ensures that the adversarial training within ReflectGAN remains grounded in realistic and environmentally consistent spectral transformations.

\subsection{Key Contributions of the ReflectGAN}

The ReflectGAN model brings several key innovations to the application of generative adversarial networks (GANs) in environmental remote sensing, particularly for transforming vegetation-contaminated satellite data into accurate bare soil reflections. Its design and implementation contribute meaningfully to both theoretical advancements and practical applications in remote sensing and land monitoring:

\begin{itemize}
    \item Conditional Spectral Translation: By conditioning on vegetated soil spectral signatures, ReflectGAN ensures that generated bare soil outputs are not only realistic but also contextually relevant, addressing a major gap in traditional GAN applications for environmental data.
    
    \item Advanced Feature Learning with Residual Blocks: The integration of residual learning within the Generator enables the model to capture complex, non-linear spectral transformations, while also mitigating challenges like vanishing gradients, leading to more stable and efficient training.
    
    \item Robust Discriminative Analysis: The tailored Discriminator architecture critically evaluates concatenated vegetated and bare soil spectra, enhancing adversarial training dynamics and improving the realism of the generated soil reflections.
    
    \item Enhanced Environmental Monitoring: ReflectGAN offers a novel approach for generating accurate soil reflectance from vegetated areas, supporting more reliable SOC estimation, soil health assessment, and agricultural monitoring from satellite imagery.
    
    \item Broader Applicability: Although designed for remote sensing, the conditional learning and spectral translation strategy of ReflectGAN can be adapted to other domains where disentangling complex mixed signals is necessary, such as urban mapping, forestry, and climate change monitoring.
\end{itemize}

Overall, ReflectGAN represents a significant advancement in the application of GANs for environmental remote sensing, combining architectural innovation with practical relevance to address longstanding challenges in soil analysis from satellite imagery.

\begin{figure*}
  \centering
  \includegraphics[width=7in]{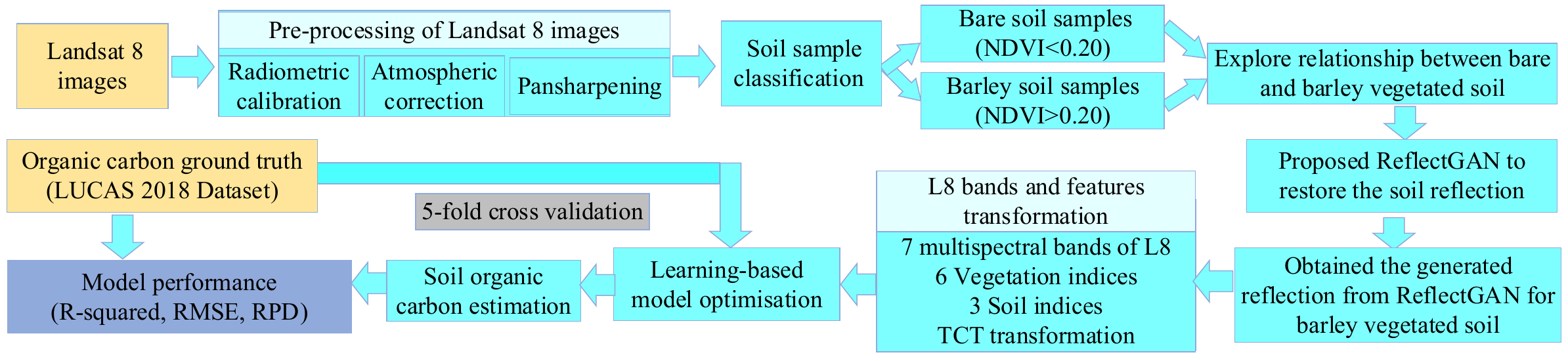}
  \caption{Illustration of the proposed framework for soil organic carbon estimation, encompassing data preprocessing, ReflectGAN-based reflectance correction, model training, and evaluation.}
  \label{Methodology}
\end{figure*}

\section{Proposed Method}

This study introduces a ReflectGAN-based framework to improve SOC estimation from L8 imagery by mitigating the impact of vegetation on spectral reflectance. The overall pipeline, illustrated in Fig.~\ref{Methodology}, begins with remote sensing data acquisition and ends with SOC estimation using both ML and DL models.

L8 imagery was preprocessed to ensure spectral consistency and spatial accuracy. This involved radiometric and atmospheric correction, followed by pansharpening using the 15-meter panchromatic band to enhance the resolution of the 30-meter multispectral bands. These steps converted raw data into surface reflectance values and enabled precise extraction of spectral signatures at soil sampling locations.

To address vegetation interference, we included both bare and vegetated soil samples, identified using the Normalized Difference Vegetation Index (NDVI). By covering NDVI values across a broader range, the study evaluates model robustness under varying vegetation conditions, especially in scenarios where vegetation masks underlying soil properties.

\begin{table}[htbp]
\centering
\caption{Summary of optimized hyperparameter configurations for each model using grid search.}
\label{hyperparameter_optimization}
\begin{tabular}{|p{0.8cm}|p{7cm}|}
\hline
\textbf{Model} & \textbf{Tuned Hyperparameters} \\
\hline
LR & Standardization: True; Random State: 42 \\
\hline
RF & Number of Estimators: 100; Random State: 42 \\
\hline
DT & Random State: 42; Max Depth: None \\
\hline
GB & Number of Estimators: 100; Random State: 42 \\
\hline
CBR & Learning Rate: 0.1; Tree Depth: 10; Loss Function: RMSE; Number of Iterations: 100 \\
\hline
KNN & Number of Neighbors: 5; Distance Metric: Euclidean \\
\hline
ANN & Hidden Layers: 3 (Units: 128-64-1); Activation: ReLU; Dropout Rate: 0.3; Batch Size: 32; Epochs: 100; Learning Rate: 0.001 \\
\hline
SVR & Kernel: RBF; Regularization (C): 1.0; Epsilon: 0.2; Batch Size: 32; Epochs: 100; Learning Rate: 0.001 \\
\hline
1dCNN & Convolutional Layers: 3; Filters: 16-32-64; Kernel Size: 3; Padding: 1; Fully Connected Layer: 128 units; Dropout Rate: 0.3; Batch Size: 32; Epochs: 100; Learning Rate: 0.001 \\
\hline
\end{tabular}
\end{table}

A core innovation in this study is the ReflectGAN model, which is specifically designed to reconstruct soil-specific reflectance from vegetation-affected L8 samples. Unlike generic image-to-image translation GANs, ReflectGAN leverages paired vegetated and bare soil samples to learn a conditional mapping that suppresses vegetation influence while preserving soil-related spectral characteristics. The generator is trained to reconstruct clean reflectance patterns that resemble bare soil, guided by a discriminator that enforces spectral realism and consistency. This design ensures that the recovered reflectance remains meaningful for downstream SOC estimation.

In addition to L8 reflectance, auxiliary features were incorporated to enhance the dataset. These included VIs (e.g., NDVI, SAVI, MSAVI), SIs (e.g., Brightness Index, Salinity Index, Dry Soil Index), and TCT components (brightness, greenness, wetness). Derived from the L8 bands, these features provide additional context on vegetation density and soil conditions relevant to SOC distribution.

The enriched dataset was used to train a range of ML and DL models, including Linear Regression (LR), Random Forest (RF), Decision Tree (DT), Gradient Boosting (GB), CatBoost Regressor (CBR), K-Nearest Neighbors (KNN), Support Vector Regression (SVR), Artificial Neural Networks (ANN), and a 1D Convolutional Neural Network (1D-CNN). Each model was systematically optimized through grid search, and the final hyperparameter settings are summarized in Table~\ref{hyperparameter_optimization}. Model performance was assessed using standard evaluation metrics: coefficient of determination ($R^2$), Root Mean Squared Error (RMSE), and Residual Prediction Deviation (RPD).

These metrics are defined as follows:

\begin{equation}
R^2 = 1 - \frac{\sum (y_{\text{true}} - \hat{y}_{\text{est}})^2}{\sum (y_{\text{true}} - \bar{y}_{\text{avg}})^2},
\end{equation}

\begin{equation}
RMSE = \sqrt{\frac{1}{N} \sum_{i=1}^{N} (y_{i,\text{true}} - \hat{y}_{i,\text{est}})^2},
\end{equation}

\begin{equation}
RPD = \frac{\text{std\_dev}}{RMSE},
\end{equation}

where $y_{\text{true}}$ and $\hat{y}_{\text{est}}$ represent the actual and estimated SOC values, $\bar{y}_{\text{avg}}$ is the mean of the actual values, $N$ is the total number of observations, and $\text{std\_dev}$ is the standard deviation of the observed values.

These metrics collectively offer insight into both estimation accuracy and generalization capability. The ReflectGAN-corrected reflectance, when used as input to these models, aims to reduce spectral noise and recover underlying soil properties, especially in vegetation-dominated areas, thereby improving SOC estimation performance.

\begin{figure}[htbp]
    \centering
    \subfigure[Lightly vegetated soil reflectance (NDVI = 0.25)]{
        \includegraphics[width=\linewidth]{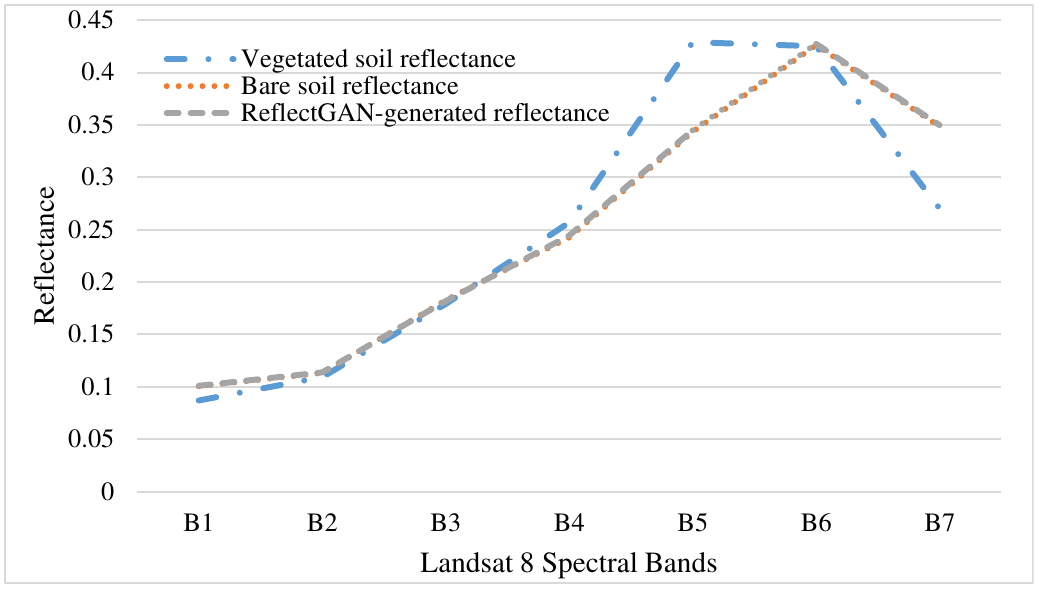}
        \label{fig:veg_reflectance}
    }
    \vfill
    \subfigure[Moderately vegetated soil reflectance (NDVI = 0.42)]{
        \includegraphics[width=\linewidth]{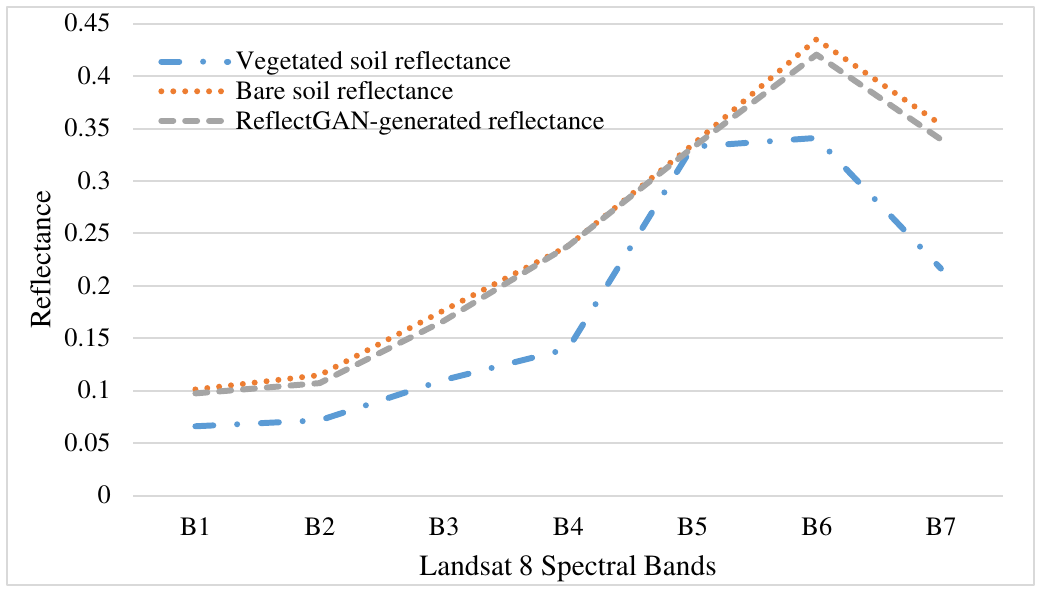}
        \label{fig:bare_reflectance}
    }
    \vfill
    \subfigure[Highly vegetated soil reflectance (NDVI = 0.83)]{
        \includegraphics[width=\linewidth]{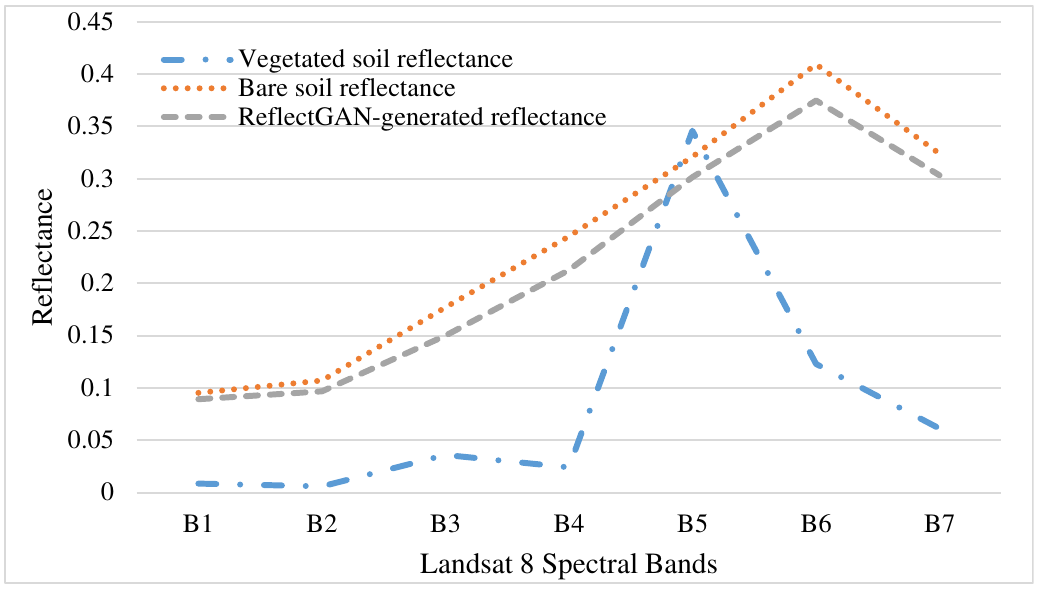}
        \label{fig:reflectgan_reflectance}
    }
    % \caption{Comparison of soil reflectance under varying vegetation densities. Each subfigure shows the vegetated soil reflectance, corresponding bare soil reflectance, and ReflectGAN-generated reflectance, illustrating the ability of ReflectGAN to reconstruct bare soil spectra across different vegetation covers.}
    \caption{Comparison of different soil reflectance types against ReflectGAN-reconstructed reflectance, illustrating how the proposed method transforms vegetated reflectance into bare soil reflectance under varying vegetation densities.}

    \label{fig:reflectance_comparison}
\end{figure}

\section{Results}

This section evaluates the effectiveness of ReflectGAN in enhancing SOC estimation under mixed bare and vegetated conditions. We first examine the correlation between spectral bands and SOC across different input types, followed by a performance comparison of ML and DL models using various reflectance scenarios.

Table~\ref{Pearson} presents Pearson's correlation coefficients between L8 spectral bands (B1–B7) and SOC for three different input types: bare soil, vegetated soil, and ReflectGAN-reconstructed reflectance. As expected, bare soil reflectance exhibited stronger negative correlations with SOC, particularly in the shortwave infrared bands (B6 and B7; –0.44 and –0.46, respectively), reaffirming their importance in SOC estimation. In contrast, vegetated samples showed substantially weaker correlations, highlighting the masking effect of vegetation on soil reflectance. Notably, ReflectGAN-reconstructed soil reflectance demonstrated the strongest correlations across all bands, especially within the red-edge and shortwave infrared regions (B4 to B7), reaching up to –0.74. 

These results indicate that ReflectGAN effectively mitigates vegetation-induced distortion and enhances the fidelity of soil reflectance, providing more reliable inputs for SOC modeling. The similarity between bare soil reflectance and ReflectGAN-reconstructed reflectance, both in terms of correlation values and visual patterns (Fig.~\ref{fig:reflectance_comparison}), indicates that ReflectGAN preserves the physical integrity of bare soil spectra even when reconstructed from vegetation-contaminated observations.

\newcolumntype{L}[1]{>{\centering\let\newline\\\arraybackslash\hspace{0pt}}m{#1}}
\newcolumntype{C}[1]{>{\centering\let\newline\\\arraybackslash\hspace{0pt}}m{#1}}
\newcolumntype{R}[1]{>{\centering\let\newline\\\arraybackslash\hspace{0pt}}m{#1}}
\newcolumntype{S}[1]{>{\centering\let\newline\\\arraybackslash\hspace{0pt}}m{#1}}

\begin{table*}[htp]
\caption{Pearson's correlation coefficients between Landsat 8 spectral bands (B1–B7) and SOC for different input types, illustrating how ReflectGAN improves spectral correlation compared to vegetated soil inputs.}

% \caption{Pearson's Correlation Coefficient Between Landsat 8 Spectral Bands and SOC for Different Input Types.}
\centering
\begin{tabular}{|L{5.5cm}|L{1.2cm}|L{1.2cm}|L{1.2cm}|L{1.2cm}|L{1.2cm}|L{1.2cm}|L{1.2cm}|}
\hline
\textbf{Input Type} & \textbf{B1} & \textbf{B2} & \textbf{B3} & \textbf{B4} & \textbf{B5} & \textbf{B6} & \textbf{B7} \\
\hline
Bare Soil Only & -0.27 & -0.29 & -0.40 & -0.36 & -0.34 & -0.44 & -0.46 \\
\hline
Vegetated Soil Only & -0.08 & -0.10 & -0.14 & -0.14 & -0.04 & -0.13 & -0.15 \\
\hline
ReflectGAN-Reconstructed Soil (Proposed) & -0.43 & -0.49 & -0.65 & -0.68 & -0.67 & -0.73 & -0.74 \\
\hline
\end{tabular}
\label{Pearson}
\end{table*}

\newcolumntype{L}[1]{>{\centering\let\newline\\\arraybackslash\hspace{0pt}}m{#1}}
\newcolumntype{C}[1]{>{\centering\let\newline\\\arraybackslash\hspace{0pt}}m{#1}}
\newcolumntype{R}[1]{>{\centering\let\newline\\\arraybackslash\hspace{0pt}}m{#1}}
\newcolumntype{S}[1]{>{\centering\let\newline\\\arraybackslash\hspace{0pt}}m{#1}}

\begin{table*}[htp]
\caption{SOC estimation performance ($R^2$\textuparrow, $RMSE$\textdownarrow, $RPD$\textuparrow) comparison of established ML and DL models using different reflectance inputs, including ReflectGAN-reconstructed reflectance.}

\centering
\begin{tabular}{|L{2.8cm}|L{1.2cm}|L{1.2cm}|L{1.2cm}|L{1.2cm}|L{1.2cm}|L{1.2cm}|L{1.2cm}|L{1.2cm}|L{1.2cm}|}
\hline
\textbf{Input} & \textbf{LR} & \textbf{RF} & \textbf{DT} & \textbf{GB} & \textbf{CBR} & \textbf{KNN} & \textbf{ANN} & \textbf{SVR} & \textbf{1D-CNN} \\
\hline
Bare Soil Only & 0.28, 9.11, 1.19 & 0.65, 6.41, 1.71 & 0.43, 7.98, 1.39 & 0.58, 6.87, 1.59 & 0.60, 6.74, 1.62 & 0.53, 7.27, 1.49 & 0.41, 8.26, 1.31 & 0.39, 8.50, 1.28 & 0.50, 7.45, 1.45 \\
\hline
Bare Soil With Additional Features & 0.29, 9.02, 1.21 & 0.61, 0.71, 1.61 & 0.31, 8.87, 1.27 & 0.55, 7.14, 1.52 & 0.64, 6.30, 1.72 & 0.53, 7.04, 1.54 & 0.44, 7.72, 1.41 & 0.40, 8.41, 1.29 & 0.51, 6.95, 1.58 \\
\hline
Vegetated Soil Only & 0.00, 6.66, 0.99 & -0.09, 6.92, 0.95 & -1.01, 9.34, 0.71 & -0.04, 6.76, 0.98 & -0.03, 6.74, 0.98 & -0.16, 7.14, 0.92 & -0.01, 6.67, 0.99 & -0.03, 6.75, 0.98 & -0.06, 6.85, 0.97 \\
\hline
Vegetated Soil With Additional Features & -0.04, 6.79, 0.98 & -0.09, 6.91, 0.96 & -0.03, 9.39, 0.70 & -0.04, 6.76, 0.98 & -0.01, 6.68, 0.99 & -0.22, 7.31, 0.90 & -0.06, 6.83, 0.97 & -0.03, 6.72, 0.98 & -0.18, 7.23, 0.92 \\
\hline
Bare Soil + Raw Vegetated & 0.17, 8.69, 1.10 & 0.55, 6.32, 1.51 & 0.25, 8.14, 1.18 & 0.47, 6.89, 1.38 & 0.51, 6.65, 1.44 & 0.39, 7.41, 1.29 & 0.34, 7.74, 1.23 & 0.22, 8.44, 1.13 & 0.47, 6.83, 1.41 \\
\hline
Bare Soil + Raw Vegetated With Additional Features & 0.15, 8.77, 1.09 & 0.52, 6.51, 1.47 & 0.24, 8.13, 1.19 & 0.48, 6.85, 1.39 & 0.50, 6.76, 1.42 & 0.39, 7.43, 1.29 & 0.34, 7.71, 1.24 & 0.25, 8.24, 1.16 & 0.40, 7.31, 1.31 \\
\hline
\textbf{ReflectGAN-Reconstructed Soil (Proposed)} & 0.50, 4.14, 1.95 & 0.53, 3.93, 2.11 & 0.27, 5.39, 1.35 & 0.50, 4.24, 1.82 & 0.53, 3.95, 2.10 & 0.47, 4.27, 1.88 & 0.50, 4.09, 2.02 & 0.54, 3.95, 2.07 & 0.51, 4.07, 2.01 \\
\hline
ReflectGAN-Reconstructed Soil With Additional Features & 0.50, 4.18, 1.91 & 0.51, 4.14, 1.94 & 0.23, 5.55, 1.30 & 0.51, 4.23, 1.82 & 0.53, 4.05, 1.96 & 0.46, 4.40, 1.78 & 0.49, 4.14, 1.98 & 0.50, 4.23, 1.86 & 0.47, 4.33, 1.83 \\
\hline
Bare Soil + ReflectGAN-Reconstructed Soil & 0.36, 7.57, 1.27 & 0.65, 5.50, 1.75 & 0.41, 7.23, 1.33 & 0.60, 5.95, 1.60 & 0.64, 5.59, 1.70 & 0.55, 6.27, 1.52 & 0.49, 6.76, 1.42 & 0.43, 7.17, 1.33 & 0.50, 6.65, 1.44 \\
\hline
Bare Soil + ReflectGAN-Reconstructed Soil With Additional Features & 0.39, 7.49, 1.29 & 0.61, 5.84, 1.64 & 0.52, 6.49, 1.47 & 0.56, 6.21, 1.53 & 0.60, 5.96, 1.61 & 0.53, 6.38, 1.51 & 0.54, 6.36, 1.51 & 0.41, 7.30, 1.31 & 0.47, 6.62, 1.49 \\
\hline
\end{tabular}

\vspace{0.2cm}
\footnotesize{
\textbf{Note:} ReflectGAN-reconstructed soil refers to reflectance reconstructed from vegetated observations using a paired GAN model specifically trained to minimize vegetation interference.
}
\label{tab:soc_estimation_results}
\end{table*}

Table~\ref{tab:soc_estimation_results} summarizes SOC estimation results across different reflectance input types. The bare soil scenario yielded strong baseline performance, with RF achieving the best results ($R^2$ = 0.65, $RMSE$ = 6.41, $RPD$ = 1.71), likely due to the absence of vegetation interference. Incorporating additional features (such as VIs, SIs, and TCT components) slightly improved performance for some models (e.g., CBR and 1dCNN), suggesting that supplementary spectral information can enhance learning in specific cases. However, for other models, including extra features led to comparable or marginally reduced performance, indicating that when vegetation effects are minimal, the benefit of additional indices remains limited.

Models trained on vegetated soil inputs exhibited poor performance across all approaches due to the dominance of plant reflectance, with several models yielding negative or near-zero $R^2$ values. The inclusion of additional features did not substantially improve these results. These findings confirm that vegetation contamination severely masks soil spectral information, and that conventional feature augmentation alone is insufficient to recover meaningful SOC-related patterns from vegetation-affected observations.

Combining vegetated and bare soil samples led to slight improvements compared to using vegetated samples alone, as the models could partially leverage information from bare soil observations. The addition of further features, such as vegetation and soil indices, resulted in only minor gains for a few models, while overall performance remained largely unchanged. Despite these adjustments, the results continued to lag behind those achieved using bare soil data alone, underscoring that vegetation contamination fundamentally limits the reliability of SOC estimation.

ReflectGAN, which reconstructs bare soil reflectance from vegetation-contaminated observations using a paired-GAN architecture, substantially improved SOC estimation performance across all models. The best result was achieved by RF, with an $R^2$ of 0.53, $RMSE$ of 3.93, and $RPD$ of 2.11, indicating strong accuracy and reliability. Other models, including LR, CBR, ANN, and 1dCNN, also achieved RPD values exceeding 2.0, demonstrating that ReflectGAN consistently enhances SOC estimation across diverse ML and DL algorithms. This general applicability underscores the robustness of ReflectGAN as a versatile solution for SOC modeling. The addition of extra features did not yield further improvements, indicating that ReflectGAN-reconstructed reflectance alone provides sufficient information for accurate SOC estimation.

Combining real bare soil samples with ReflectGAN-reconstructed reflectance resulted in the highest $R^2$ values across several models, with RF achieving the best $R^2$ of 0.65. However, RMSE and RPD values were not consistently optimal compared to using ReflectGAN-reconstructed data alone, likely due to distributional differences between real and generated reflectance. The addition of extra features led to minor improvements, particularly in models such as LR, DT, and ANN, while most other models showed comparable or slightly reduced performance. These results suggest that although hybrid training helps capture broader spectral variability, it may also introduce additional complexity that limits further gains in estimation reliability.

The results demonstrate that models trained on bare soil reflectance consistently achieve superior SOC estimation performance compared to those using vegetated inputs. ReflectGAN-reconstructed reflectance further enhanced model accuracy by minimizing vegetation-induced distortions, enabling reliable SOC estimation across mixed land cover conditions. These findings highlight the potential of reflectance reconstruction strategies to improve soil monitoring across heterogeneous landscapes.

\section{Discussion}

This section explores the performance and broader significance of the proposed ReflectGAN framework for SOC estimation. We first compare ReflectGAN against existing vegetation correction methods, then evaluate its generalizability across different satellite datasets. Finally, we discuss its practical strengths and limitations, propose future research directions, and highlight the broader implications for operational soil monitoring and remote sensing applications.

\subsection{Comparison with Existing Vegetation Correction Methods}

We evaluate the performance of the proposed ReflectGAN approach against a range of existing vegetation correction methods, encompassing both traditional techniques and advanced DL-based models from recent literature. The selected models reflect widely used or representative approaches in the literature that address vegetation interference via index-based filtering, physical modeling, spectral unmixing (linear and probabilistic), and GAN-based translation. We excluded more recent nonlinear GAN-based unmixing models such as HyperGAN \cite{wang2024pixel} and 3D-GAN \cite{suresh2023spectral} because they are primarily designed for abundance estimation and spectral classification rather than full reflectance reconstruction, which is essential for SOC estimation in our application. The comparative results, summarized in Table~\ref{Comparison}, provide insights into the effectiveness of each method in supporting SOC estimation under vegetation-influenced conditions.

The RF model was chosen for this comparison as it consistently performed well in most cases (Table~\ref{tab:soc_estimation_results}). Additionally, RF is a robust and widely used algorithm for SOC estimation due to its simplicity and effective handling of high-dimensional data.

\begin{table*}[htp]
\caption{Performance comparison of the proposed ReflectGAN method against existing vegetation correction approaches for SOC estimation using the best-performing RF model, with uniform hyperparameters applied across all methods to ensure fair evaluation.}

\centering
\begin{tabular}{| L{3.2cm} | C{3.2cm} | C{1.5cm} | C{1.5cm} | C{1.5cm} |}
\hline
\textbf{Approach} & \textbf{Vegetation Handling} & \textbf{$R^2$} & \textbf{RMSE} & \textbf{RPD} \\
\hline
Bare Soil Only & No Vegetation Correction (Bare Soil Samples Only) & 0.65 & 6.41 & 1.71 \\
\hline
Vegetation Index-Based Correction \cite{xue2017significant} & Partial Correction (NDVI, SAVI) & -0.09 & 6.90 & 0.95 \\
\hline
Spectral Mixture Analysis (SMA) \cite{veraverbeke2012spectral, salih2023spectral} & Linear Spectral Mixing Model & 0.21 & 8.25 & 1.14 \\
\hline
Adaptive Spectral Mixture Analysis (ASMA) \cite{sun2023global} & Adaptive Linear Spectral Unmixing & 0.11 & 6.38 & 1.06 \\
\hline
Radiative Transfer Model (PROSAIL) \cite{verhoef2007coupled} & Physically Based Model Correction & 0.24 & 8.13 & 1.19 \\
\hline
Probabilistic Mixture Model-Based Spectral Unmixing (PMM-SU) \cite{eches2010bayesian, hoidn2024probabilistic} & Bayesian Spectral Unmixing & 0.40 & 6.95 & 1.45 \\
\hline
Pix2Pix GAN \cite{isola2017image} & Paired Image-to-Image Translation & 0.16 & 6.16 & 1.09 \\
\hline
CycleGAN \cite{zhu2017unpaired} & Unpaired Image-to-Image Translation & 0.08 & 7.10 & 1.02 \\
\hline
\textbf{ReflectGAN (Proposed)} & \textbf{Soil-Specific GAN-Based Transformation} & \textbf{0.53} & \textbf{3.93} & \textbf{2.11} \\
\hline
\end{tabular}
\label{Comparison}
\end{table*}

Bare soil only serves as a baseline scenario where only samples with no or minimal vegetation cover were used. As expected, this approach achieved the highest performance among all conventional methods ($R^2$ = 0.65, RMSE = 6.41, RPD = 1.71), primarily due to the absence of vegetation interference, which enabled a direct relationship between soil reflectance and SOC to be captured. However, despite its strong performance, this approach has limited practical scalability, as bare soil pixels are infrequent and seasonally constrained across most agricultural and natural landscapes.

While bare soil samples provide the most reliable inputs, many landscapes are dominated by partial or full vegetation cover, necessitating correction strategies. Vegetation Index-Based Correction methods, such as NDVI and SAVI, were tested to mitigate vegetation influence; however, they showed poor performance ($R^2$ = –0.09, RMSE = 6.90, RPD = 0.95). These indices indirectly account for vegetation cover but fail to recover the true underlying soil reflectance. As a result, they are ineffective at disentangling mixed signals, particularly when vegetation density varies across samples.

SMA moderately improved performance ($R^2$ = 0.21, RMSE = 8.25, RPD = 1.14), indicating some success in linearly separating vegetation and soil components. However, it assumes fixed endmembers and linear mixing, assumptions that often do not hold under real-world conditions where non-linear reflectance interactions are common, limiting its effectiveness.

ASMA achieved a slightly better RMSE than SMA but resulted in a lower $R^2$ (0.11), likely due to unstable estimation of endmember fractions. While ASMA adaptively adjusts spectral endmembers for each pixel, this flexibility can introduce noise, particularly under dense vegetation cover.

PROSAIL, a physically based correction model, yielded marginal improvements over SMA ($R^2$ = 0.24) but still exhibited a high RMSE. Although PROSAIL accurately simulates canopy–soil reflectance interactions, its reliance on numerous input parameters (e.g., LAI, leaf angle, soil moisture) limits its practical applicability across large, heterogeneous landscapes.

PMM-SU achieved improved performance ($R^2$ = 0.40, RPD = 1.45) by leveraging Bayesian inference to robustly estimate soil and vegetation fractions. Its probabilistic framework accounts for uncertainty, offering better generalization than linear SMA approaches; however, it still struggles to recover true bare-soil reflectance under high vegetation densities.

Among DL-based approaches, Pix2Pix GAN demonstrated modest gains ($R^2$ = 0.16), but struggled to generalize beyond paired training data. Although it outperformed traditional unmixing methods, its outputs occasionally retained vegetation artifacts, limiting its effectiveness for accurate SOC estimation.

CycleGAN, trained without paired supervision, performed the worst among GAN-based models ($R^2$ = 0.08), likely due to misalignment between input and target domains. Without explicit one-to-one correspondence between vegetated and bare soil samples, CycleGAN struggled to learn consistent spectral transformations under conditions of high spectral variability.

Finally, ReflectGAN (Proposed) significantly outperformed all competing approaches ($R^2$ = 0.53, RMSE = 3.93, RPD = 2.11). By learning soil-specific spectral transformations from paired vegetated and bare soil reflectance samples, ReflectGAN effectively suppresses vegetation-induced noise and reconstructs physically realistic soil signals. Its superior performance highlights the critical importance of domain-specialized generative models for addressing complex reflectance mixing challenges in SOC estimation.

To rigorously assess the differences among the vegetation correction approaches, we conducted a one-way Analysis of Variance (ANOVA) separately for each SOC estimation metric ($R^2$, RMSE, and RPD) based on cross-validation results. The ANOVA indicated statistically significant differences across methods for all three metrics ($p$ < 0.001). Subsequently, Tukey’s Honest Significant Difference (HSD) post-hoc analysis was performed to identify specific pairwise comparisons. The results confirmed that ReflectGAN significantly outperformed all other approaches, exhibiting higher $R^2$ and RPD values and lower RMSE values across all comparisons (all pairwise $p$ < 0.01). These findings provide strong statistical evidence supporting the superiority of ReflectGAN for SOC estimation relative to both traditional vegetation correction methods and baseline GAN architectures.

In summary, traditional vegetation correction techniques are often constrained by oversimplified assumptions and limited spectral separation capabilities. In contrast, GAN-based methods—particularly ReflectGAN—demonstrate strong potential for advancing SOC estimation by learning complex, non-linear transformations that more accurately reconstruct true soil reflectance under vegetative conditions.

\subsection{Evaluating ReflectGAN Performance on Sentinel-2 Data}

\newcolumntype{L}[1]{>{\centering\let\newline\\\arraybackslash\hspace{0pt}}m{#1}}
\newcolumntype{C}[1]{>{\centering\let\newline\\\arraybackslash\hspace{0pt}}m{#1}}
\newcolumntype{R}[1]{>{\centering\let\newline\\\arraybackslash\hspace{0pt}}m{#1}}
\newcolumntype{S}[1]{>{\centering\let\newline\\\arraybackslash\hspace{0pt}}m{#1}}

\begin{table*}[htp]
\caption{Descriptive Statistical Parameters for the SOC Dataset Investigated Using Sentinel-2 Data}
\centering
\begin{tabular}{| L{2.8cm} | L{2.8cm} | L{1.2cm} | L{1.2cm} | L{1.2cm} | L{1.2cm} | L{1.2cm} | L{1.2cm} | L{1.2cm} |}
\hline
\textbf{Soil Type} & \textbf{NDVI Range} & \textbf{Sample No.} & \textbf{Min} & \textbf{Max} & \textbf{Mean} & \textbf{Median} & \textbf{Std. Dev.} & \textbf{CV(\%)} \\
\hline
Mixed & $0 < \text{NDVI} < 0.93$ & 229 & 2.7 & 81.4 & 17.35 & 14.80 & 10.43 & 60.17 \\
\hline
\end{tabular}
\label{dataset_statistic_S2}
\end{table*}

\newcolumntype{L}[1]{>{\centering\let\newline\\\arraybackslash\hspace{0pt}}m{#1}}
\newcolumntype{C}[1]{>{\centering\let\newline\\\arraybackslash\hspace{0pt}}m{#1}}
\newcolumntype{R}[1]{>{\centering\let\newline\\\arraybackslash\hspace{0pt}}m{#1}}
\newcolumntype{S}[1]{>{\centering\let\newline\\\arraybackslash\hspace{0pt}}m{#1}}

\begin{table*}[htp]
\caption{SOC Estimation Performance ($R^2$\textuparrow, $RMSE$\textdownarrow, $RPD$\textuparrow) Using ML and DL Models with Sentinel-2 Reflectance Inputs.}
\centering
\begin{tabular}{| L{2.8cm} | L{1.2cm} | L{1.2cm} | L{1.2cm} | L{1.2cm} | L{1.2cm} | L{1.2cm} | L{1.2cm} | L{1.2cm} | L{1.2cm} |}
\hline
\textbf{Input} & \textbf{LR} & \textbf{RF} & \textbf{DT} & \textbf{GB} & \textbf{CBR} & \textbf{KNN} & \textbf{ANN} & \textbf{SVR} & \textbf{1D-CNN} \\
\hline
Bare Soil Only & 0.40, 6.11, 1.33 & 0.63, 4.80, 1.70 & 0.27, 6.45, 1.25 & 0.59, 5.02, 1.62 & 0.64, 4.78, 1.74 & 0.61, 4.96, 1.66 & 0.52, 5.38, 1.55 & 0.48, 6.00, 1.49 & 0.59, 5.04, 1.63 \\
\hline
Vegetated Soil Only & 0.03, 19.95, 1.02 & -0.11, 21.16, 0.97 & -1.41, 30.05, 0.70 & -0.07, 20.99, 0.99 & 0.00, 20.29, 1.01 & -0.01, 20.41, 1.01 & 0.06, 19.62, 1.05 & -0.01, 20.70, 0.99 & 0.04, 19.91, 1.04 \\
\hline
Bare Soil + Raw Vegetated Soil & -0.01, 15.11, 1.02 & 0.04, 14.52, 1.09 & -1.05, 21.02, 0.76 & 0.03, 14.90, 1.11 & 0.10, 14.50, 1.10 & -0.03, 14.95, 1.05 & 0.03, 14.75, 1.05 & 0.08, 15.10, 1.04 & 0.00, 14.84, 1.04 \\
\hline
\textbf{ReflectGAN-Reconstructed Soil (Proposed)} & 0.32, 7.25, 1.50 & \textbf{0.55,
5.35, 2.26} & 0.44, 6.07, 2.11 & 0.55, 5.66, 2.00 & 0.56, 5.53, 2.09 & 0.54, 5.88, 1.91 & 0.49, 6.13, 1.81 & 0.36, 7.66, 1.46 & 0.53, 5.73, 1.99 \\
\hline
Bare Soil + ReflectGAN-Reconstructed Soil & 0.04, 9.29, 1.02 & 0.62, 5.56, 1.72 & 0.47, 6.57, 1.46 & 0.53, 6.29, 1.53 & 0.59, 5.87, 1.65 & 0.61, 5.85, 1.62 & 0.20, 9.20, 1.25 & 0.30, 9.31, 1.90 & 0.44, 8.26, 1.34 \\
\hline
\end{tabular}
\label{S2}
\end{table*}

To validate the generalizability of ReflectGAN beyond L8 data, additional experiments were conducted using S2 multispectral imagery. The corresponding SOC dataset, summarized in Table~\ref{dataset_statistic_S2}, encompasses a wide range of SOC values (2.7–81.4 g/kg) and vegetation conditions (NDVI ranging from 0 to 0.93), providing a robust and diverse testbed for performance evaluation.

As shown in Table~\ref{S2}, models trained exclusively on bare soil samples achieved strong performance, with CBR yielding the highest results ($R^2$ = 0.64, RMSE = 4.78, RPD = 1.74). In contrast, models trained directly on vegetated soil samples without any correction exhibited severe performance degradation across all approaches, underscoring the significant spectral distortion introduced by vegetation cover. Simply merging bare and vegetated samples without correction also led to poor performance, further highlighting the challenges of mixed soil–vegetation reflectance.

Training models on ReflectGAN-corrected reflectance substantially improved SOC estimation. Among the tested models, RF achieved the best results using corrected data ($R^2$ = 0.55, RMSE = 5.35, RPD = 2.26), confirming the effectiveness of ReflectGAN in recovering soil-specific information from vegetated observations. Furthermore, combining real bare soil samples with ReflectGAN-corrected data enhanced model robustness, with RF achieving strong estimation accuracy ($R^2$ = 0.62, RMSE = 5.56, RPD = 1.72).

Overall, these findings demonstrate that ReflectGAN effectively mitigates vegetation-induced spectral contamination and enables accurate SOC mapping across heterogeneous landscapes containing both bare and vegetated surfaces, where traditional approaches typically struggle.

\subsection{Limitations and Future Work}

While this study highlights the effectiveness of ReflectGAN in improving SOC estimation through vegetation correction of satellite-derived reflectance, several important limitations should be recognized.

First, ReflectGAN relies on a paired dataset of vegetated and corresponding bare soil reflectance, which may not always be feasible. In this study, barley-vegetated samples were statistically matched with their nearest bare soil counterparts based on spatial proximity and spectral consistency. While this pairing strategy improves data reliability in controlled agricultural settings, it may limit scalability in heterogeneous landscapes where suitable bare references are sparse. Future research could investigate unpaired or semi-supervised GAN frameworks—such as domain adaptation GANs—to relax this requirement and improve model adaptability. Additionally, extending the approach to other types of vegetation would further test its robustness across diverse land cover conditions.

Second, the current model was trained and evaluated within a limited geographic scope and soil diversity. Although generalization was assessed using S2 imagery, broader validation across diverse agroecological zones is necessary to confirm its robustness. Incorporating additional environmental variables such as soil moisture, texture, and seasonal dynamics could further improve the estimation accuracy and operational flexibility of the model.

Third, while ReflectGAN improves the correlation between corrected reflectance and SOC, the reconstructed spectra may not always perfectly represent physically realistic soil characteristics. Integrating prior knowledge or physics-based constraints, such as hybrid physics-aware neural networks, could lead to more interpretable and scientifically consistent outputs. Moreover, future extensions could explore explainable DL frameworks and uncertainty-aware correction mechanisms to further enhance model transparency for operational soil monitoring.

Finally, this study treated reflectance correction as a standalone preprocessing step for SOC estimation. Future research could explore end-to-end DL architectures that jointly optimize both reflectance correction and SOC estimation within a unified framework. Additionally, incorporating temporal satellite observations and vegetation phenology information may further improve the robustness and reliability of SOC estimation across dynamic agricultural landscapes.

\subsection{Practical Implications for Remote Sensing and Soil Monitoring}

ReflectGAN offers notable practical advantages for large-scale soil monitoring, remote sensing, and agricultural applications. By reconstructing bare-soil reflectance from vegetation-contaminated satellite observations, ReflectGAN reduces dependence on seasonally scarce bare-soil pixels, thereby expanding the temporal and spatial coverage of SOC estimation efforts. This capability is particularly valuable in regions with persistent vegetation cover, where traditional SOC mapping approaches often fail or require extensive masking.

Beyond coverage improvements, ReflectGAN is also computationally efficient. Compared to conventional GAN-based architectures such as Pix2Pix and CycleGAN, ReflectGAN employs a streamlined generator–discriminator design with fewer parameters, leading to faster convergence during training and reduced computational overhead. Although traditional spectral unmixing or physical models such as SMA, ASMA, and PROSAIL are lightweight per sample, they often demand extensive ancillary data, manual parameter tuning, or rely on rigid assumptions that limit large-scale scalability. In contrast, ReflectGAN operates efficiently and scalably across diverse regions using only remote sensing reflectance data, without requiring additional field measurements or complex initialization procedures. This enables seamless integration into operational remote sensing workflows supporting precision agriculture, sustainable land management, and carbon monitoring initiatives.

By enabling robust SOC estimation from mixed bare and vegetated surfaces, ReflectGAN helps address key challenges in remote sensing-based soil assessment, providing new opportunities for continuous, scalable soil monitoring across dynamic agricultural and natural landscapes.

\section{Conclusion}

This study introduced ReflectGAN, a novel paired-GAN framework specifically designed to reconstruct bare-soil reflectance from vegetation-contaminated satellite observations. By explicitly learning the transformation between vegetated and bare-soil reflectance, ReflectGAN effectively mitigates vegetation-induced distortion, providing cleaner inputs for SOC estimation. Extensive experiments on both L8 and S2 datasets demonstrated that ReflectGAN consistently outperformed traditional vegetation correction methods and existing GAN-based models, achieving significantly improved SOC estimation accuracy. Beyond its methodological contributions, ReflectGAN offers a scalable and computationally efficient solution for enhancing soil property mapping in vegetation-rich landscapes, paving the way for more reliable remote sensing applications in agriculture, land management, and environmental monitoring.

% or


% \appendices

% % you can choose not to have a title for an appendix
% % if you want by leaving the argument blank
% \section{}

\section*{Data and Code Availability}
The source code for the ReflectGAN model and the paired dataset (L8) used in this study are publicly available at: \url{https://github.com/DristiDatta/ReflectGAN}.

\section*{Acknowledgment}
This work has been supported by the Cooperative Research Centre for High Performance Soils whose activities are funded by the Australian Government's Cooperative Research Centre Program.

%\bibliographystyle{IEEEtran}
%\bibliography{bibtex/bib/IEEEexample}

\ifCLASSOPTIONcaptionsoff
  \newpage
\fi

\end{document}